\documentclass[%
 reprint,
 superscriptaddress,
 groupedaddress,
 nofootinbib,
 amsmath,amssymb,
 aps,
 pra,
 floatfix,
]{revtex4-2}

\usepackage{amsfonts}
\usepackage{amsthm}
\usepackage{mathtools}
\usepackage{physics}
\usepackage{siunitx}
\usepackage{graphicx}
\usepackage{bm} 
\usepackage{bbold} 
\usepackage{microtype}
\usepackage{enumerate}
\usepackage{xifthen}
\usepackage[hyperfootnotes=true,linkcolor=blue,menucolor=black,colorlinks=false]{hyperref}
\usepackage[capitalise]{cleveref}
\usepackage[acronym]{glossaries}
\glsdisablehyper
\usepackage{dsfont}

\newacronym{mm}{MM}{metastable manifold}
\newacronym{ems}{EMS}{extreme metastable states}
\newacronym{ann}{ANN}{Artificial neural network}
\newacronym{hnn}{HNN}{Hopfield Neural Network}
\newacronym{am}{AM}{Associative Memory}

\newcommand{\disp}[1]{\mathcal{D}[#1]}
\newcommand{\Q}{\mathcal{Q}}
\newcommand{\Id}{\mathds{I}}
\newcommand{\opa}{\hat{a}}
\newcommand{\opad}{\hat{a}^\dagger}
\newcommand{\opn}{\hat{n}}
\newcommand{\lv}{\mathcal{L}}

\newcommand{\hilbert}{\mathcal{H}}

\newcommand{\R}{\mathbb{R}}
\newcommand{\half}{\frac{1}{2}}

\newcommand{\inpar}[1]{\left( #1 \right)}
\newcommand{\insqr}[1]{\left[ #1 \right]}
\newcommand{\inbrc}[1]{\left\lbrace #1 \right\rbrace}

\newenvironment{eqs}[1][]
  {\subequations\ifthenelse{\isempty{#1}}{}{\label{#1}}\align}
  {\endalign\endsubequations}

\newcommand{\ccite}[1]{Ref.~\cite{#1}}

\DeclarePairedDelimiter{\floor}{\lfloor}{\rfloor}
\graphicspath{{pfigs/}}

\begin{document}

\title{Quantum memories for squeezed and coherent superpositions in a driven-dissipative nonlinear oscillator}

\author{Adrià Labay-Mora}
\email{alabay@ifisc.uib-csic.es}
\author{Roberta Zambrini}
\author{Gian Luca Giorgi}%
\affiliation{%
 Institute for Cross-Disciplinary Physics and Complex Systems (IFISC) UIB-CSIC, Campus Universitat Illes Balears, Palma de Mallorca, Spain.
}%

\date{\today}

\begin{abstract}
    Quantum oscillators with nonlinear driving and dissipative terms have gained significant attention due to their ability to stabilize cat states for universal quantum computation. Recently, superconducting circuits have been employed to realize such long-lived qubits stored in coherent states. We present a generalization of these oscillators, which are not limited to coherent states. The key ingredient lies in the presence of different nonlinearities in driving and dissipation, beyond the quadratic one. Through an extensive analysis of the asymptotic dynamical features for different nonlinearities, we identify the conditions for the storage and retrieval of quantum states, such as squeezed states, in both coherent and incoherent superpositions. We explore their applications in quantum computing, where squeezing prolongs the lifetime of memory storage for qubits encoded in the superposition of two symmetric squeezed states, and in quantum associative memory, which has so far been limited to the storage of classical patterns.
\end{abstract}

\maketitle

\section{Introduction}
Quantum oscillators with high nonlinearities have recently gained attention due to their promise to perform universal quantum computation \cite{Terhal_2020,cai2021reviewQEC,mirrahimi2014universal,lescanne2020exponential}. These types of systems benefit from an infinite dimensional Hilbert space, which allows, for instance, autonomous quantum error correction techniques \cite{Gertler2021,leghtas2013hardware,lieu2020symmetryQEC} and fault-tolerant quantum computation \cite{puri2019faultolerant,ofek2016breakevenQEC,xu2023autonomous}. Experimental realizations of nonlinear dissipative oscillators have also been carried out in the last decade by several groups that were able to engineer up to five-photon dissipation using supercomputing resonators \cite{grimm2020kerrstabilization,svensson2017three,svensson2018multiple,chang2020three}, including demonstrations of logical qubits encoded in oscillators of this kind \cite{heeres2017implementing,xu2020demonstration,berdou2022one}. Moreover, there have been proposals to use such oscillators as a resource for quantum machine learning algorithms where information might be encoded in the amplitude or phase of a squeezed state \cite{zhong2020boson,madsen2022boson}. Examples include quantum reservoir computing \cite{mujal2021opportunities,govia2021qrc} and quantum associative memory \cite{labay2022memory}. 

In all the cases studied so far, the exchange of photons with the environment -- in the form of dissipation -- and the nonlinear driving, have been considered to involve the same non-linearity, i.e. photon processes up to the same degree (where $n$-photon driving and dissipation are balanced). In this regime, it is well known that the ground state (in the case of Kerr oscillators) \cite{grimm2020kerrstabilization,mishra2010kerrsqueezing} or the steady state (in the case of dissipative oscillators) \cite{mirrahimi2014universal} is a cat-like superposition or classical mixture (in the presence of single-photon loss) of coherent states \cite{cai2021reviewQEC,lang2021multi6}. These coherences that appear in the steady state can be traced back to the symmetry of the system which can be weak or strong depending on the system parameters \cite{minganti2023dissipative}.

In this work, we explore different photon number exchange processes, yielding squeezed states with distinct properties. Our findings show (i) an extended metastable phase, particularly for two symmetrically squeezed states, enabling prolonged qubit memory, and (ii) the system's capability to generate genuine quantum patterns in quantum associative memory, highlighting its versatility.

The first main result (i) builds on previous work where bit flips are exponentially suppressed while phase flips increase only linearly in qubits encoded in a superposition of coherent states\cite{lescanne2020exponential,berdou2022one}. Similar to other works where squeezed-cat states can enhance the storage time compared to coherent-cat states \cite{jeannic2018slow,minganti2022squeezedcat,hillmann2023squeezedcat,xu2023autonomous}, the use of squeezed states preserves the scaling of the two characteristic timescales of the memory when the nonlinear degrees are not coprime. The numerical results demonstrate that they can reduce the error rate and thus, extend the lifetime of quantum information. 

The second main result of this paper (ii) extends our previous work in \ccite{labay2022memory} where we showed that such systems can be used as a quantum associative memory algorithm, as they permit the retrieval of previously stored patterns in the form of coherent states. Here, the novelty is the use of genuine quantum patterns encoded in the amplitude and phase of squeezed states. In fact, our platform enables the storage and retrieval of truly quantum memories, which is unprecedented in the existing literature~\cite{Gopalakrishnan,diamantini2006quantum,PhysRevLett.107.277201,PhysRevA.95.032310,rotondo2018open,PhysRevA.98.042308,carollo2021exactness,fiorelli2021phase,enhancing2021marsh}. We notice also that the lower discrimination of squeezed states for small mean photon numbers reduces the storage capacity so that the coherent state case is optimal under this lens.

The article is organized as follows. In \Cref{sec:model} we introduce the master equation and review some of its properties. In \Cref{sec:symmetry} we determine the type of symmetry (weak or strong) of the system depending on the (non)linearity of coherent and dissipative terms, and in \Cref{sec:metastability} we study the metastable phase that arises in the case of weak symmetry. The following \Cref{sec:squeezing} is devoted to characterizing the squeezed states that form the metastable manifold of the system. All this analysis allows us to explore two applications for these oscillators in \Cref{sec:memory} and \Cref{sec:qam}. In the former, we explore the capability of the system to store quantum information over time to be used in quantum computation \cite{mirrahimi2014universal}. In the latter, we extend the proposal introduced in \ccite{labay2022memory} to implement a quantum associative memory for pattern discrimination. We summarize the main results of the paper in \Cref{sec:discussion}.

\section{The model}\label{sec:model}

\begin{figure}
    \centering
    \includegraphics[width=\linewidth]{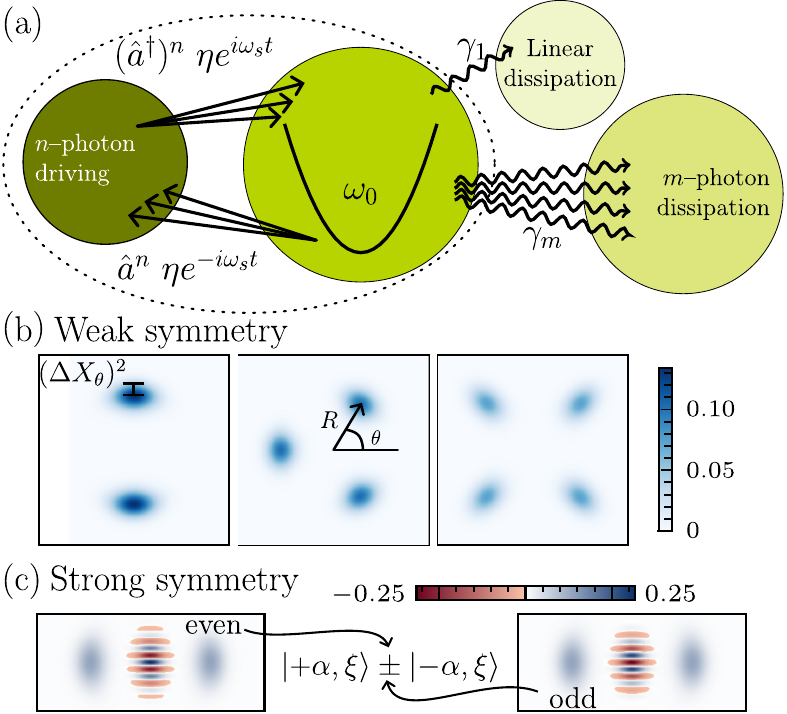}
    \caption{(a) Sketch of the driven-dissipative nonlinear oscillator with the three processes involved in the master equation: nonlinear periodic driving with degree $n$, linear dissipation with rate $\gamma_1$ and nonlinear dissipation of degree $m$. The driving force pushes the system with strength $\eta$ and frequency $\omega_s$ that may deviate from the natural oscillator frequency $\omega_0$. The dissipative terms emit photons out of the system at rates $\gamma_1$ and $\gamma_m$ for the single- and multiple-photon processes respectively. (b) Wigner distribution of the steady states generated in the weak-symmetry regime with $\gamma_1 > 0$. From left to right: $(n,m) = \{ (2,3), (3,4), (4,3) \}$. (c) Wigner distribution of the two steady states (corresponding to even- and odd-parity eigenstates) present in the strong-symmetry regime with $\gamma_1 = 0$ and $(n,m) = (2,4)$.}
    \label{fig:sketch}
\end{figure}

The system under study consists of a generalized driven-dissipative nonlinear oscillator described by the Gorini-Kossakowski-Lindblad-Sudarshan (GKLS) master equation introduced in \ccite{labay2022memory}
\begin{equation} \label{eq:pp_me_nm}
    \pdv{\rho}{t} = -i [H_n, \rho] + \gamma_1 \disp{\opa}\rho + \gamma_m \disp{\opa^m} \rho\ \equiv \lv \rho\ ,
\end{equation}
where in the Liouvillan superoperator $\lv$ we distinguish three different terms. First, the unitary evolution described by the Hamiltonian, which in the rotation frame and after the parametric approximation is
\begin{equation} \label{eq:pp_ham_nm}
    H_n = \Delta \opad\opa + i \eta_n \left[\opa^n e^{i \theta_0 n} - (\opad)^n e^{-i \theta_0 n} \right]\ .
\end{equation}
This models an $n$-photon drive with $ n \ge 1$, where the detuning between the natural oscillator frequency $\omega_0$ and the frequency of the driving force $\omega_s$ is defined as $\Delta = \omega_0 - \omega_s$. This $n$-photon parametric process produces squeezing effects for $n> 1$ \cite{braunstein1987gensqueezing} and will be called the squeezing term in the following. The parameter $\eta_n$ controls the driving strength and $\phi$ represents its phase.

The system is also coupled to the environment through two Lindblad dissipators $\disp{\hat{O}}\rho = \hat{O}\rho \hat{O}^\dagger - (1/2)\{\hat{O}^\dagger\hat{O},\rho\}$. The first is the unavoidable linear (single-photon) dissipation typical of oscillators of this type, $\hat O=\hat a$, characterized by a photon-loss rate $\gamma_1$ \cite{Carmichael1993Open}. The second is an engineered nonlinear term with $m$-photon exchange rate $\gamma_m$ ($m > 1$), dissipating photons to the environment in groups of $m$. A sketch of the different nonlinear processes can be seen in \cref{fig:sketch}(a). We note that other sources of dissipation such as dephasing terms have been excluded from the analysis.

This type of resonator has been extensively studied in the literature when $n = m$ \cite{gevorkyan1999coherent,mundhada2017dissipation,mirrahimi2014universal,berdou2022one}. Generalizations with standard and higher order Kerr terms have also been reported \cite{minganti2023critical,minganti2023dissipative}. Furthermore, these terms in Hamiltonians may arise from static nonlinearities as in squeezed Kerr resonators \cite{venkatraman2022static,frattini2022squeezed} or from lower order coupled modes \cite{albert2019pair} creating pair-cat codes. Moreover, superconducting circuits can be used to engineer $n$-photon driving and dissipation terms using a single buffer drive by modifying the flux frequency going through a Josephson junction \cite{svensson2017three,svensson2018multiple,chang2020three}. While a single buffer drive can only implement a photon exchange mechanism of a particular degree $n$, it may be possible to couple the resonator to two buffer drives, each with $n$ and $m$-photon driving and dissipation terms. Then, by increasing the strength of the desired degree and decreasing the others, one recovers \cref{eq:pp_me_nm}.

Oscillators of this kind have recently been proposed to implement universal quantum computation and fault-tolerant quantum computation due to their capacity to autonomously protect the qubits from bit-flip errors \cite{mirrahimi2014universal,cai2021reviewQEC,lieu2020symmetryQEC,leghtas2013hardware,Gertler2021}. Although phase flips still need to be corrected from photon loss, they can be prevented using quantum error correction techniques. The presence of only $n$-photon-exchange processes allows only coherent states to be stored. In the following sections we will see that the resonator described by \cref{eq:pp_me_nm} can be used to store squeezed states as well as coherent superpositions of such states. In particular, the type of state that is preserved depends on the symmetry of the system.

In general, the results presented in the following have been obtained by setting the parameters $\gamma_1 = 1.0$, $\gamma_m = 0.2$, and $\Delta = 0.4$, unless specified otherwise. Also, to simplify the notation, we will typically refer to the master equation with $n$-photon driving and $m$-photon dissipation as a pair of numbers $(n,m)$.

\section{Symmetry and steady-state structure} \label{sec:symmetry}

In this section, we discuss the role of symmetry for \cref{eq:pp_me_nm} and its implications for understanding and controlling its dynamics. We explore how symmetry can be used to design robust quantum systems and protect against decoherence and other forms of environmental noise \cite{lieu2020symmetryQEC,lihm2018quditqed}. We remark that although the system is infinite-dimensional, a cutoff of the dimension is introduced to perform the numerical simulation. This approximation guarantees the presence of at least one steady state by Evan's theorem~\cite{evans1977generators,baumgartner2008analysis}. Also note that the existence of at least one steady state for $n = m$ has been shown analytically for infinite dimension\cite{azouit2015convergence,azouit2016well}.

In accordance with \ccite{minganti2023dissipative}, the system exhibits a $\mathds{Z}_n$ symmetry when $n = m$, with the symmetry being weak (strong) when $\gamma_1 > 0$ ($\gamma_1 = 0$). This argument can be generalized for $n \neq m$ where, however, the absence of linear dissipation alone does not guarantee the presence of a strong symmetry, for which a necessary condition is that the Hamiltonian and all the jump operators commute with $\hat{Z}_p = \exp(-i \pi \opad \opa / p)$ \cite{albert2014symmetries,marco2020symmetry}. Hence,
\begin{equation}
    [\hat{Z}_p, \opa^n] = [\hat{Z}_p, \opa^m] = 0 \Rightarrow p = \gcd(n,m) > 1\ ,
\end{equation}
where $p$ determines the number of steady states in the system. Notably, instances of strong symmetry arise when the dissipation degree, $m$, is not a coprime of $n$, as exemplified by cases such as $(2,4)$, $(3, 6)$, or $(4, 6)$. On the other hand, when the two powers are coprime $\gcd(n,m) = 1$, there is a weak symmetry $\mathds{Z}_n$ such that $[\lv, \mathcal{Z}_n] = 0$, where $\mathcal{Z}_n[\bullet ]=e^{-i\pi\hat{a}^\dagger\hat{a}/n}\bullet e^{i\pi\hat{a}^\dagger\hat{a}/n}$.

The symmetry allows us to block-diagonalize the Liouvillian by dividing its matrix into $n$  ($p^2$) independent sectors in the case of weak  (strong) symmetry. Specifically, for weak symmetry, the Liouvillian can be expressed as $L = \bigoplus_{\mu=0}^{n-1} B_\mu^W$, where the steady state is located within the symmetry sector $\mu = 0$ \cite{minganti2018spectral,albert2014symmetries}. Conversely, for strong symmetry, the Liouvillian takes the form $L = \bigoplus_{\mu,\nu=0}^{p-1} B_{\mu\nu}^S$. The $p$ steady states are found in the sectors $B_{\mu\mu}^S$ with $\mu=0,\dots,p-1$. The other sectors contain the coherences that will eventually decay in the long-time limit. 

In \cref{fig:sketch}, we present illustrative examples of the different steady states produced by \cref{eq:pp_me_nm}. These have been obtained numerically by solving the steady-state equation $\lv \rho_{\mathrm{ss}} = 0$ \cite{numerics}. The Wigner distribution of these states allows us to observe non-classical phenomena as indicated by the negativity of this quasi-probability distribution \cite{kenfack2004negativity}. On the one hand, \cref{fig:sketch}(b) shows various scenarios with weak symmetry, characterized by a single steady state. Notably, we observe variations in the shape of the lobes corresponding to different photon exchange powers.  On the other hand, \cref{fig:sketch}(c) exhibits the case of $(2, 4)$ under strong symmetry, which leads to two steady states in the system. These steady states correspond to even and odd parity cat-states with squeezed states as lobes. A similar situation can be expected for $(3, 6)$, where we anticipate three steady states with well-defined symmetry eigenvalues $\mu=0,1,2$. We note that in these two cases $\gcd(n,m) = n$ resulting in the number of steady states being equal to the power of the driving. However, special situations arise when $\gcd(n,m) = p \neq n$, with $p > 1$. For example, in the case of $(4,6)$, although there are four symmetrically distributed squeezed lobes, there are only two steady states due to $p = 2$. Similar to $(2,4)$, these two steady states also have well-defined parity with $\mu=0,1$. This particular case will be discussed in more detail in \cref{apx:4n6m}.

In both cases, for weak and strong symmetry, we can approximate the lobes as squeezed-coherent states
\begin{equation} \label{eq:squeezed_state}
    \ket{\alpha,\xi} = D(\alpha) S(\xi) \ket{0}
\end{equation}
where $D(\alpha) = \exp(\alpha \opad - \alpha^* \opa)$ is the displacement operator with amplitude $\alpha = r\exp(i \theta)\in \mathds{C}$. Here, $r \in [0,\infty)$ determines the distance from the origin and $\theta \in [0, 2\pi)$ the displacement angle in phase-space. Then, $S(\xi) = \exp\{-(\xi (\opad)^2 - \xi^* \opa^2)\}$ is the squeezing operator with squeezing parameter $\xi = s\exp(i\phi) \in \mathds{C}$. The magnitude $s\in\R$ determines the strength of the squeezing while $\phi \in [0,\pi)$ is the direction in which these states have a squeezed quadrature. We can relate the squeezing strength $s$ to the variance of the quadrature as
\begin{equation} \label{eq:dx_from_s}
    \ev{(\Delta X_\phi)^2} = \frac{1}{4} e^{-2 s}
\end{equation}
where $X_\phi = [\opa \exp(-i \phi) + \opad \exp(i \phi)] / 2$. 

By analyzing the mean field equation (detailed in \cref{apx:mean_field}), we can determine the phase of each lobe in phase space (corresponding to the coherent displacement), given by $\theta_j = \theta_0 + (2j + 1) / n$ for $j=1,\dots,n$. Consequently, the amplitudes of the $n$ lobes composing the steady state are ${ \alpha_j = r\exp(i \theta_j)}_{j=1}^n$, where $r$ is approximated by \cref{eq:mf_amplitude}. A mean-field approximation does not capture squeezed fluctuations, but looking at the direction of the smallest quadrature of the lobes in \cref{fig:sketch}(b), we appreciate that when $n > m$ the states are phase squeezed, and when $n < m$ the states are amplitude squeezed. Thus, taking into account the phase of the lobes $\theta_j$, we get 
\begin{equation} \label{eq:lobe_xi}
    \xi_j = s \exp[i 2\theta_j + i \Theta(n - m) \pi/2]\qquad j =1,\dots,n
\end{equation}
where $\Theta(x)$ is the Heaviside function which is one for $x > 0$ and vanishes otherwise. The case $n =m$ is the previously studied case with no squeezing ($s=0$) \cite{labay2022memory}. Here, we assume $r$ and $s$ are the same for each lobe due to the rotational symmetry of the system. 

The steady state of the system in the presence of weak symmetry is then a classical mixed state of the lobes so
\begin{equation} \label{eq:steadystate_weak}
    \rho_{\mathrm{ss}}^W = \frac{1}{n} \sum_{j=1}^n \op{\alpha_j,\xi_j}\ .
\end{equation}
In contrast, when the system has strong symmetry, we are left with $p$ steady states which are coherent superpositions of the lobes. An example of such a state is the one belonging to the $\mu=0$ symmetry sector
\begin{equation} \label{eq:steadystate_strong}
    \rho_{\mathrm{ss}}^S = \op{\psi_{\mathrm{ss}}^S};\qquad \ket{\psi_{\mathrm{ss}}^S} = \frac{1}{\sqrt{n}} \sum_{j=1}^n \ket{\alpha_j,\xi_j} \ .
\end{equation}

Throughout the rest of the article, we work mainly in the weak symmetry regime with $\gamma_1 > 0$ which corresponds to the most physical scenario.

\section{Metastability}\label{sec:metastability}

For a system described by the GKLS master equation $\partial_t \rho = \lv \rho$ \cite{lindblad1976generators,gorini1976completely}, the dynamics can be understood in terms of the set of complex eigenvalues $\inbrc{\lambda_j}$ of the (non-Hermitian) Liouvillian superoperator $\lv$ and of the right ($\inbrc{R_j}$) and left ($\inbrc{L_j}$) eigenvectors, obeying $\lv R_j = \lambda_j R_j$ and $\lv^\dagger L_j = \lambda_j^* L_j$, respectively, with normalization $\tr(L_j^\dagger R_k) = \delta_{jk}$ \cite{minganti2018spectral}. Then, assuming the presence of at least one steady state $\rho_{\mathrm{ss}}$ with $\lambda_1 = 0$ (which is always true in finite dimensions \cite{evans1977generators,baumgartner2008analysis}), the time evolution of a state $\rho(0)$ can be decomposed as
\begin{equation} \label{eq:pp_modes_decomp}
    \rho(t) = \rho_{\mathrm{ss}} + \sum_{j > 1} \tr[L_j^\dagger \rho(0)] e^{\lambda_j t} R_j\ ,
\end{equation}
where for convenience the eigenvalues are sorted such that $0 \ge \Re\lambda_j \ge \Re\lambda_{j + 1}$.

As discussed in \cite{macieszczak2016towards,minganti2018spectral}, in open quantum systems, metastability typically arises as a consequence of a separation between two consecutive eigenvalues of the Liouvillian, i.e. $\tau_l \gg \tau_{l+1}$ where $\tau_l^{-1} = - \Re \lambda_l$ and it is closely related to the emergence of quantum entrainment and dissipative phase transitions \cite{cabot2021metastable}. In the presence of metastability, fast dynamics ($t < \tau_{l+1}$) is well separated from  the slow metastable phase ($t > \tau_{l+1}$). After the relatively fast transient $\tau_{l+1}$, the system decays into the metastable manifold whose size depends on the number of slowly decaying modes. The emergence of a metastable phase, which isolates $l$ modes from the rest, enables us to confine the system dynamics within this metastable manifold spanned by the right eigenmodes $\{ R_j \}_{j=1}^l$ of the Liouvillian \cite{cabot2021metastable,macieszczak2021metastability}. Specifically, after the initial decay ($t > \tau_{l+1}$), the system's state can be expressed as a complex linear combination of only the aforementioned eigenmodes. Finally, a decay to the steady state occurs for times $t > \tau_2$.

Metastability is seen in the Liouvillian of \cref{eq:pp_me_nm} when there is linear dissipation (weak symmetry) and the nonlinear powers are balanced ($n=m$) \cite{cabot2021metastable,labay2022memory,minganti2023dissipative}. In the following we remove the latter constraint while staying in the weak symmetry limit ($\gamma_1 > 0$). Given the large number of parameters and regimes of the model, for the sake of clarity, we consider Liouvillians with at most four-photon driving exchange ($n$) and we modify the power of the dissipation ($m$) to be above and below that of the driving. Moreover, we compare with the coherent-state situation ($n=m$) to study in which situation squeezing can improve the performance of certain applications.

 One of the first properties to analyze is the length of the metastable phase. For that, we recall that in the balanced situation the separation in the Liouvillian spectrum occurs between the eigenvalues $n$ and $n+1$, where $n$ is the degree of the driving term 
  \cite{labay2022memory}. Furthermore, the separation between the two eigenvalues increases for large $\eta_n$ and small $\gamma_n$. Both of these properties remain valid in the unbalanced situation $n \neq m$ but the scaling of the separation with respect to the ratio $\eta_n/\gamma_m$ is different for each case (see \cref{apx:mean_field}). Consequently, to compare the different cases, we use the parameter $\ev{\opn}_{\mathrm{ss}} = \tr (\opn \rho_{\mathrm{ss}})$ corresponding to the mean photon number of the $n$ lobes forming the steady state [\cref{eq:steadystate_weak}].

\begin{figure}
    \centering
    \includegraphics[width=\linewidth]{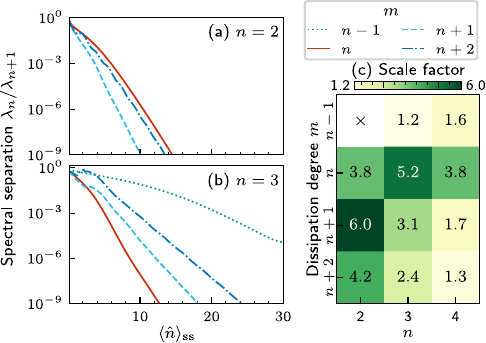}
    \caption{ Ratio between the eigenvalues $\lambda_n$ and $\lambda_{n+1}$ defining the Liouvillian separation for (a) $n=2$ and (b) $n=3$. Line styles corresponding to $m$ values in the top right inset. (c) Scale factor $k$ (see main text) of the spectral separation with the photon number, obtained by fitting the lines in (a-c) to an exponential function.}
    \label{fig:spectral_separation_234}
\end{figure}

In \cref{fig:spectral_separation_234}(a,b), we show the ratio between the real parts of the eigenvalues $\lambda_n$ and $\lambda_{n + 1}$ which relates to the Liouvillian separation. The smaller the ratio, the longer the metastable phase. In this sense, we see the ratio decreasing exponentially with the photon number, as $\Re \lambda_n \to 0^-$ and $\Re \lambda_{n+1} \to -\infty$. The behavior is the same for all powers (including $n = 4$, which is not shown), but the slope of the curves changes drastically. In general, for $n > 2$, having different power between the driving and the dissipation reduces the eigenvalue separation, leading to a shorter metastable phase. This is not the case for $n = 2$, where both situations with $m > n$ ($m = 3$ and $m = 4$) result in a larger separation for the same mean photon number.

The slope of the curves can be used to better compare the different scenarios. For that, we calculate the scaling factor $k$ of the separation with the mean photon number in \cref{fig:spectral_separation_234}(c). This has been obtained from an exponential fit of the lines in the left panels as $10^{b \ev{\opn}_{\mathrm{ss}}} a$ where $a$ and $b$ are the fit parameters that relate to the scale factor as $k = 10^b$. We can see that squeezing in the presence of two lobes improves the metastable time. The worst scaling occurs for $n = 4$ and $m=6$ which is related to the different distribution of the Liouvillian spectra. More details can be found in \cref{apx:4n6m}. Also, although the case $(2,1)$ is shown in this type of figure, it has not been studied since the single steady state is a squeezed-vacuum state with no metastability.

The appearance of the metastable phase pauses the dynamics of the system which can be described by only the slowest $n$ modes. However, the eigenmodes $\{ R_j \}_{j=1}^{n}$ themselves are not valid quantum states since they are traceless. The quantum states that span the metastable manifold are known as extreme metastable states which we will denote as $\{ \mu_j \}_{j=1}^n$. They can be constructed using the extreme eigenvalues of the left eigenmodes $\{ L_j \}_{j=1}^n$ as $\mu_j = \sum_{a=1}^n c_{a}^{M,m} R_a$ where $c_{a}^{M}$ and $c_a^m$ are the maximum and minimum eigenvalues of $L_a$. Of course, the contribution of the steady state ($a=1$) is maximal since $L_1 = \Id$ so $c_1^M = c_1^m = 1$. This gives the unit trace condition to the extreme metastable states. The other coefficients can be chosen as those that minimize the classicality condition \cite{macieszczak2021metastability}. Notably, we found that the combination of eigenmodes is the same for all values $m$ (for a fixed driving degree $n$). The actual expression can be seen in the Supplemental Material of \ccite{labay2022memory}.

Using this method, we isolate the $n$ lobes that form the steady state (see \cref{fig:sketch}) which correspond identically to the $n$ extreme metastable states. In this way, the steady state can be reconstructed as $\rho_{\mathrm{ss}}^W = (1/n)\sum_{j} \mu_j$, like in \cref{eq:steadystate_weak}.

\section{Characterization of squeezed states} \label{sec:squeezing}

\begin{figure*}
    \centering
    \includegraphics[width=\linewidth]{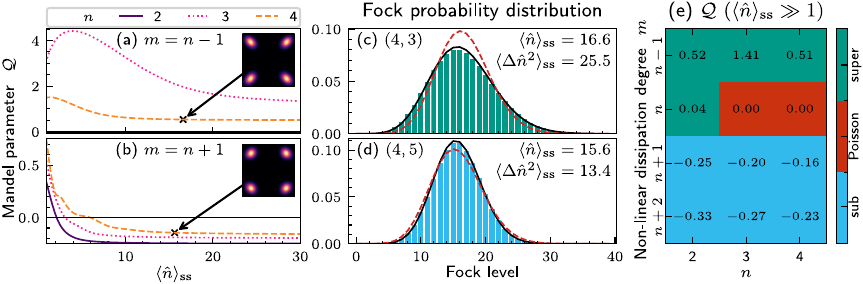}
    \caption{(a) and (b) Evolution of the Mandel $\Q$ parameter for increasing value of the average photon number. We distinguish between dissipative powers below the driving degree (a) and above (b). (c) and (d) Probability distribution of the steady state marked in crosses on the respective left figure. The solid black line corresponds to the distribution of a steady state as a mixture of $n$ squeezed states. The dashed red line corresponds to the probability distribution of a coherent state with the mean photon number shown. The super- and the sub-Poissonian character of the states are identified given the larger and smaller variance of the bars in (c) and (d) respectively. (e) Mandel $\Q$ parameter for large average photon number $\ev{\opn}_{\mathrm{ss}} \gg 1$.}
    \label{fig:complete_figure_q}
\end{figure*}

We have seen in \cref{sec:symmetry} that the steady state of the oscillator is formed by $n$ symmetrically distributed lobes which are, depending on the symmetry, entangled (strong) or mixed (weak). In both cases, the number of lobes is determined by the squeezing degree $n$. The power of the dissipation, however, modifies the shape of such states. While it is well known that coherent states can be obtained for equal nonlinear powers ($n = m$) \cite{rotondo2018open,gevorkyan1999coherent}, for $n \neq m$, the states become squeezed. Squeezing is a well-known quantum phenomenon in which quantum states have quantum fluctuations below the shot noise level of coherent states in one quadrature of the field. It is related to sub-Poissonian statistics, characterized by Mandel's $\Q$ parameter~\cite{mandel1979sub,mandel_wolf_1995}
\begin{equation}
    \Q = \frac{\ev{(\Delta\opn)^2} - \ev{\opn}}{\ev{\opn}}.
\end{equation}
Quantum states can be classified into sub-Poissonian ($-1 \le \Q < 0$) and super-Poissonian ($\Q > 0$). Coherent states have $\Q = 0$ since the photon number distribution follows a Poisson distribution with a mean photon number equal to their variance.

By considering the steady state of the system obtained numerically by solving $\lv \rho_{\mathrm{ss}} = 0$ in the weak symmetry regime ($\gamma_1 = 1$), we proceed to evaluate Mandel's parameter. It is noteworthy that we can directly compute $\Q$ using the steady state itself. This holds because the operator $\opn$ commutes with the symmetry operator $\hat{Z}_n$.

In \cref{fig:complete_figure_q} we study the cases where the dissipation degree $m$ is one below [\cref{fig:complete_figure_q}(a) and \cref{fig:complete_figure_q}(c)] or above [\cref{fig:complete_figure_q}(c) and \cref{fig:complete_figure_q}(d)] the driving power $n$. First, in panels (a) and (b), we plot the evolution of $\Q$ as the mean photon number of the steady state increases. We recall that the mean photon number is related to the ratio $\eta_n/\gamma_m$ as obtained from the mean-field analysis (\cref{eq:mf_amplitude}). More details on this can be found in \cref{apx:mean_field}. 

These figures illustrate two distinct scenarios: for $m = n - 1$ the states exhibit super-Poissonian behavior, while for $m = n + 1$ the states display sub-Poissonian statistics (for $\ev{\opn}_{\mathrm{ss}} > 5$). This distinction becomes evident when examining panels (c) and (d), which depict the photon number probability distribution for the steady states represented in the inset figures (a) and (b), respectively. The probability distribution of a coherent state with the same mean photon number $\ev{\opn}_{\mathrm{ss}}$ as the squeezed state is also included as a red-dashed line for comparison. We find that the states arising from $m = n + 1$ correspond to amplitude-squeezed states, characterized by sub-Poissonian statistics. Importantly, this behavior extends to other values of $m$ where $m > n$. For the case $m = n - 1$, however, we cannot draw definitive conclusions about the classicality of the states. It is the analysis of the Wigner distribution that shows that they exhibit phase squeezing, suggesting the need for alternative techniques to accurately determine their nature.

In \cref{fig:complete_figure_q}(e), we present the values of the Mandel parameter attained for large mean photon numbers ($\ev{\opn}_{\mathrm{ss}} \gg 1$). These values are computed as the average of $\Q$ when $\ev{\opn}_{\mathrm{ss}} \in [20, 30]$\footnote{The variance of this mean vanishes in most cases since $\Q$ is constant for these large values of the mean photon number.}. Additionally, we include the corresponding values for the case of $n = m$, representing coherent states, and $m = n + 2$, which also leads to sub-Poissonian statistics. Studying the figure, we can discern that the degree of sub-Poissonianity diminishes as $n$ increases, while it intensifies with growing $m$.

\begin{figure}
    \centering
    \includegraphics[width=\linewidth]{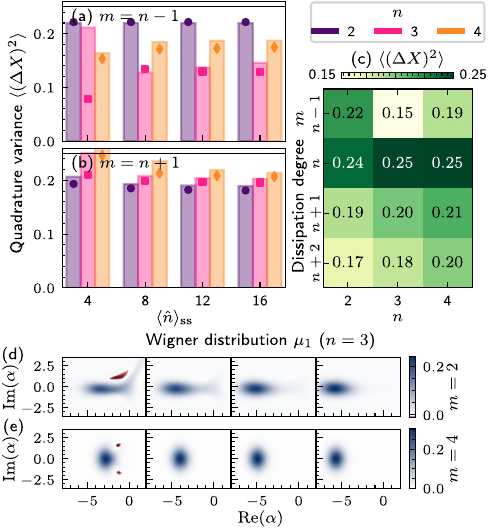}
    \caption{Quadrature variance of the extreme metastable states for the phase (a) and amplitude (b) squeezed states for $m=n-1$ and $m=n+1$, respectively. The squeezing angle corresponds to the phase of each lobe in phase space, with an extra phase factor of $\pi/2$ for amplitude squeezed states. The markers show the quadrature squeezing obtained from fitting the numerical steady-state with \cref{eq:steadystate_weak}. (c) Quadrature variance $\ev{(\Delta X)^2}$ on average for large  $\ev{\opn}_{\mathrm{ss}} \in [20, 30]$. (d,e) Wigner representation of the metastable state $\mu_1$ ($\theta_1 = \pi$) for $n = 3$. The mean photon number of the lobes, from left to right, corresponds to $\ev{\opn}_{\mathrm{ss}} = 4, 8, 12, 16$. We appreciate (d) the phase squeezing and (e) the amplitude squeezing, which leads to the phase factor $\pi/2$.}
    \label{fig:quadratures}
\end{figure}


Although the Mandel $\Q$ parameter allows us to distinguish between states with sub- and super-Poissonian statistics, it does not provide information about the extent of squeezing in a direction different from the amplitude. Concretely, for the phase-squeezed states we encounter when $n > m$, this needs to be extracted from the variance of the quadrature operator \cref{eq:dx_from_s}. The angle $\phi$ is given by the direction of minimal squeezing which can be related to the angle of the lobes in phase space as in \cref{eq:lobe_xi}.

To analyze the squeezing properties of the metastable states, we computed the variance of the quadrature operator in the direction of minimal squeezing for each state. The obtained values, presented in \cref{fig:quadratures}(a), represent the average variance over all lobes. It is important to note that, due to the rotational symmetry of the system, the quadrature variance and the associated squeezing parameter ($s$) are the same for all lobes (up to numerical precision). Moreover, as the lobes are in general mixed states which may deviate from pure coherent squeezed states, we performed a fitting procedure to determine the parameter $s$ by matching the steady states with the mixed state given in \cref{eq:steadystate_weak}. The results of the fitting process are depicted as markers in \cref{fig:quadratures}(a). This fitting approach helps determine the nature of the extreme metastable phases since a lower variance than for coherent states in a given direction does not necessarily indicate that the states are squeezed coherent states. In this way, we can assess how closely the numerical steady states resemble the expected mixed-state superposition of symmetrically distributed squeezed coherent states in \cref{eq:squeezed_state}.

In all the cases examined, we find that the quadrature variance is smaller than that of coherent states [$\ev{(\Delta X_{\mathrm{coh}})^2} = 0.25$]. The lowest quadrature variance is observed in the case $(3,2)$. However, it should be noted that the metastability window is practically non-existent for small values of $\ev{\opn}_{\mathrm{ss}}$ (refer to \cref{fig:spectral_separation_234}). In this regime, the deviations from classical behavior are significant, indicating that the dynamics cannot be adequately described by considering only the first $n$ modes. Concretely, in \cref{fig:quadratures}(d) we show the Wigner distribution of the extreme metastable state corresponding to $\theta_1 = \pi$, constructed using the extreme eigenvalues as explained. This state, for a small photon number, contains negativities and lacks a well-defined direction of squeezing. For larger amplitudes, it becomes possible to accurately construct the lobes and calculate the quadrature variance, which reaches a value of 0.15 (corresponding to $\SI{-2.21}{dB}$). Additionally, \cref{fig:quadratures}(e) shows the complementary scenario $(3,4)$. The Wigner distributions for various mean photon numbers ($\ev{\opn}_{\mathrm{ss}} = 4, 8, 12$ and $16$) are displayed. Unlike the previous case, the lobe can already be discerned for the smallest photon number. However, a significant discrepancy is observed when comparing the quadrature fluctuations obtained from the lobe [represented by the pink bar in \cref{fig:quadratures}(b)] with those obtained using the fitting procedure (indicated by the square marker). This discrepancy suggests that the state is not accurately described by \cref{eq:squeezed_state} until $\ev{\opn}_{\mathrm{ss}} \ge 8$. In the other cases, the fit reaches a close value when compared to the direct computation of the quadrature fluctuations. Thus, the validity of the approximation of the metastable states by pure squeezed coherent states is proved.

Similar to our observation in \cref{fig:complete_figure_q}, where the Mandel parameter reaches a stable value for large photon numbers, we find that the quadrature variance also stabilizes as the photon number increases. We plot this convergence value in \cref{fig:quadratures}(c), obtained by averaging $\ev{(\Delta X)^2}$ for $\ev{\opn}_{\mathrm{ss}} = 20,\dots,30$. As expected, coherent states are only obtained when $n=m$, indicating a balance between driving and dissipation. However, we can consistently generate squeezed states when the driving and dissipation degrees differ. Furthermore, in this limit of large photon number, it is worth noting that the magnitude of the squeezing is solely determined by the powers $n$ and $m$ and remains invariant under changes in the other oscillator parameters ($\gamma_1$ or $\Delta$).

\section{Memory lifetime}\label{sec:memory}

In this section, we investigate the feasibility of using the resonator as a storage medium for squeezed and cat states. We focus again on the case of weak symmetry, commonly observed in experiments due to linear losses of bosonic systems. In this regime, as mentioned before, the stationary state is an incoherent superposition of lobes, but cat states can still be displayed in the metastable transient allowing for efficient quantum state storage. We focus on evaluating two key properties: the relaxation and dephasing time. These measures are commonly employed in quantum computation to assess the memory's ability to retain information over time \cite{mirrahimi2014universal,ofek2016breakevenQEC,cai2021reviewQEC,berdou2022one}. The relaxation (or bit-flip) time is the time it takes for a lobe to decay to the steady state, while the dephasing (or phase-flip) time is the time it takes for coherences to vanish. 

In the context of bosonic memories, the bit-flip time is defined as the decay time required for a lobe $\ket{\psi_k}$ to lose all information about its initial state, resulting in the system reaching the fully mixed state $\Id = \rho_{\mathrm{ss}} = (1/n)\sum_k \op{\psi_k}$. Here we assume that the extreme metastable phases form an $n$-dimensional computational basis. For coherent states ($n=m=2$), experimental studies have demonstrated an exponential increase in the bit-flip time as a function of the mean photon number. Conversely, the phase-flip error rate exhibits a linear dependence on $2\ev{\hat{n}} / T_1$, where $T_1$ is the resonator lifetime~\cite{lescanne2020exponential,berdou2022one}.

For the quadratic driving, the quantities evaluated in the following sections are equivalent to the bit-flip and phase-flip times studied in the literature. In particular, the computational basis ($Z$ eigenvectors) consists of the lobes $\inbrc{\ket{+\alpha, \xi}, \ket{-\alpha,\xi}}$ and the even (odd) parity cat states correspond to the positive (negative) eigenvector of the $X$ operator\footnote{There might be some discrepancy compared to some works where the Bloch sphere rotates around $Y$ axis so cat-states are the logical qubits instead of the lobes themselves \cite{puri2019faultolerant,grimm2020kerrstabilization}.}. For higher driving degrees, we will straightforwardly extend the meaning of these two quantities to qudits without imposing a particular qubit encoding in each case (see \ccite{mirrahimi2014universal} for mappings taking states generated with $n=4$ onto a two-dimensional Bloch sphere). While better encodings may lead to longer storage times, we want to compare the possible advantage of using squeezed states over coherent states with the same nonlinearity in the driving. This can already be seen in the simple scenario as more complex encodings are still limited to the relaxation of the lobes. The only exception is the case $(4,6)$ due to the special symmetry of the system, as we will see.

\paragraph{Relaxation time}
To calculate the relaxation time, we compute the full master equation evolution for each state in the computational basis, i.e. $\{ \mu_k \}_{k=1}^n$. At each time step, we measure the expectation value of $\opa$, which asymptotically approaches zero as $\Tr(\opa \rho_{\mathrm{ss}}) = 0$. Thus, the relaxation time, denoted by $T_{\mathrm{rel}}$, is obtained by fitting the imaginary part\footnote{We could also use the real part or the absolute value of the operator.} of $\ev{\opa}$ to an exponentially decaying function $\exp(-T_{\mathrm{rel}} t)$. In most cases, this decay time is expected to be determined by the Liouvillian spectral gap, denoted by $\tau_2^{-1} = -\Re \lambda_2$, which represents the decay time of the slowest eigenmode.\footnote{We note that, in general, and in contrast with our model, the knowledge of the Liouvillian eigenvalues may not be sufficient to determine the time scales of a system \cite{haga2021liouvillian,mori2021metastability,flynn2021topology}.}

To explore the scaling behavior with increasing lobe separation, we perform this fitting procedure for several values of the mean photon number $\ev{\opn}_{\mathrm{ss}}$. In our analysis, we fix the dimension of the Hilbert space to $\dim{\hilbert} = 50$ to ensure an accurate representation of the dynamics over a range of mean photon numbers from $2$ to $20$. 

\begin{figure}
    \centering
    \includegraphics[width=\linewidth]{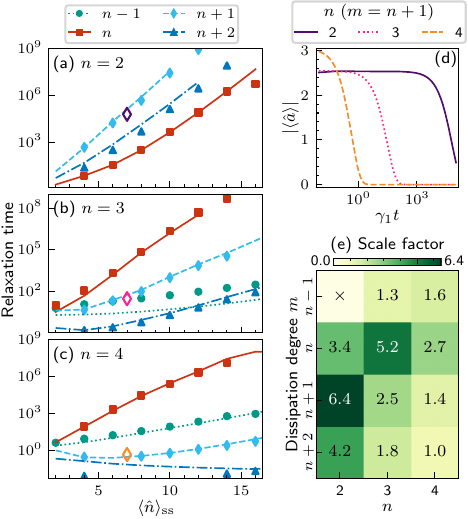}
    \caption{(a)-(c) Logarithmic plot of the relaxation time ($\gamma_1 T_{\mathrm{rel}}$) over the mean photon number of the steady state for (a) $n=2$, (b) $n=3$ and (c) $n=4$. The markers are obtained by fitting the time evolution of $\Im \ev{\opa}$ to an exponentially decaying function (error bars are smaller than marker size). The lines correspond to the decay time of the spectral gap $\lambda_2$. (d) Example full master equation evolution of $\Im \ev{\opa}$ ($y$ axis) for an initial state corresponding to the lobe $\ket{\psi_0}$ with $\ev{\opn} = 9$ ($\dim{\hilbert} = 50$). Each line corresponds to a different driving degree $n$ where we fixed the relation with the dissipation degree to $m = n + 1$. We obtain relaxation times of $\num{1.41e5}$, $70$ and $1.1$ (units of $\gamma_1$) for $n=2,3$ and $4$, respectively. (e) Scale factor of the relaxation with respect to the number of photons, obtained by fitting the data in (a)-(c) to an exponential function.}
    \label{fig:bitflip}
\end{figure}

The results are presented in \cref{fig:bitflip}, where panels (a)-(c) depict the relaxation time $T_{\mathrm{rel}}$ as a function of the mean photon number of the steady state. Each data point represents a fitting procedure applied to the decay time of $\ev{\opa}$ obtained from a full master equation evolution, using the lobes as initial states. Additionally, \cref{fig:bitflip}(d) showcases three specific trajectories with $\ev{\opn} = 9$ and different driving degrees ($n=2, 3, 4$), resulting in relaxation times on the order of $10^5$, $10^2$, and $1$ respectively. It is important to note that the error in estimating the relaxation time is less than $10^{-3}$ in all cases, with a correlation coefficient between the data and the fitting function $r^2 \sim 1 - 10^{-6}$. 

In these figures, we also include the decay of the spectral gap in lines of different strokes for each value of $m$ (dotted for $n-1$, solid for $n$, dashed for $n+1$, and dash-dotted for $n+2$). We appreciate that the relaxation time obtained from the full master equation evolution is equivalent to the decay time of the spectral gap. Hence, the evolution can be fully understood from the Liouvillian eigenvalues. 

We note that in most cases $T_{\mathrm{rel}}$ grows exponentially with the mean photon number. This behavior is well known and has been experimentally demonstrated for coherent states ($n=m=2$) \cite{lescanne2020exponential}, and we show that it remains valid for squeezed states with $n\neq m$.  In \cref{fig:bitflip}(e) we can see the scale factor $K$ of $T_{\mathrm{rel}}$ with the mean photon number obtained by exponentially fitting the data points in the left panels to $T_{\mathrm{rel}} = x K^{\ev{\opn}_{\mathrm{ss}}}$. Notably, for $n=2$, the presence of squeezing can significantly enhance the bit-flip time of the resonator, with a scale factor of $K = 6.4$ for $m=3$. This is in agreement with previous results where squeezing can enhance the storage time of a qubit \cite{jeannic2018slow,minganti2022squeezedcat,hillmann2023squeezedcat}. In general, however, squeezing is counterproductive in both amplitude and phase. The longer relaxation time for $n > 2$ is obtained when both nonlinearities are equal. The significant reduction in storage time observed is a consequence of the lobes becoming increasingly indistinguishable as their number grows. In other words, a higher photon number is required to resolve the lobes with the same precision as in the case of $n = 2$. This effect is further amplified by the presence of amplitude squeezing, which diminishes the separation between the lobes.


\paragraph{Dephasing time}
We proceed to evaluate the dephasing time, which quantifies the duration for a superposed state to lose its coherences. In our analysis, we adopt an approach similar to that for the relaxation time. We consider the $n$-cat states $\{|C_\mu^{(n)}\rangle\}_{\mu=1}^n$, derived from the $n$ lobes, as our initial states. These $n$-cat states exhibit well-defined parities, denoted by $p$, which correspond to the symmetry eigenvalues associated with each state. By investigating the expectation value over time of the projector $\hat{P}_\mu =\sum_{a=0}^{\floor{D/n}} \op{an + \mu}$ associated with the symmetry sector $p$, we observe that the expectation value progressively decays. Eventually, when the probability of finding the state within that sector reaches $1/n$ (fully mixed states), all coherences are lost. The fitting of this decay to $\langle\hat{P}_\mu\rangle = [(n-1) \exp(- \Gamma_{\mathrm{dep}} t) + 1] / n$ determines the dephasing error rate $\Gamma_{\mathrm{dep}}$.

\begin{figure}
    \centering
    \includegraphics[width=\linewidth]{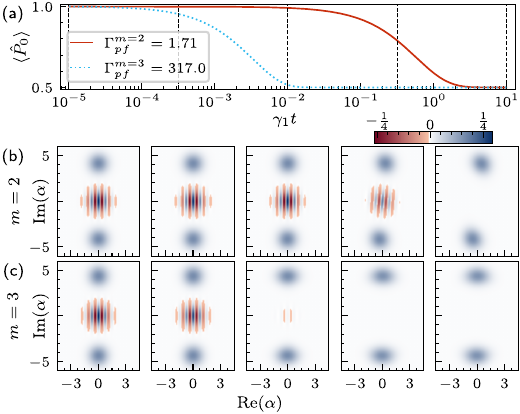}
    \caption{(a) Full master equation evolution of the parity operator for a cat-state with even parity and $\ev{\opn} = 9$ in a resonator with driving degree $n = 2$ and dissipation degree $m = 2$ (red solid line) and $m = 3$ (blue dotted line). The phase flip error rate calculated by fitting the lines to an exponentially decaying function is shown in the legend. (b) and (c) Wigner distribution of the states at times corresponding, from left to right, to the vertical dashed lines in (a), for (b) $m=2$ and (c) $m=3$.}
    \label{fig:phaseflip_evolution}
\end{figure}

An example can be seen in \cref{fig:phaseflip_evolution}(a) for $n= 2$ where we take as the initial state the even cat-state $\ket{C_0} = (\ket{\alpha} + \ket{-\alpha})/\sqrt{2}$ that populates only the even energy levels of the oscillator. Then, the evolution of the even parity operator $\hat{P}_0$ indicates the decay of the coherence, which is lost at times $0.01\gamma_1$ and $1\gamma_1$ for $m = 3$ and $2$, respectively. The Wigner representation of the states at particular times is shown in panels (b) and (c) for the two dissipation degrees. The presence of coherence can be appreciated in the negative fringes in the center of the Wigner distribution which decay faster in the presence of squeezing. 

The values of the phase-flip error rate $\Gamma_{\mathrm{dep}} = 1 / T_{\mathrm{dep}}$ for the two trajectories in \cref{fig:phaseflip_evolution}(a) are $\Gamma_{\mathrm{dep}}^{m=2} = (1.71472 \pm 0.00016)\gamma_1$ and $\Gamma_{\mathrm{dep}}^{m=3} = (316.82 \pm 0.20)\gamma_1$ so the cat state stored in the balanced situation remains in memory longer than in the unbalanced case.

\begin{figure}
    \centering
    \includegraphics[width=\linewidth]{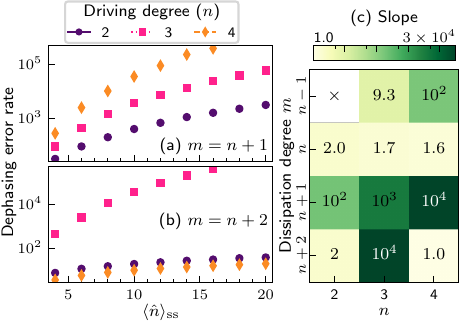}
    \caption{Dephasing error rate $\Gamma_{\mathrm{dep}}/\gamma_1$ as a function of the mean photon number for $n=2,3$, and $4$ and (a) $m=n+1$ and (b) $m=n+2$. The markers are obtained by fitting the decay of the corresponding block observable $\hat{P}_\mu$ to an exponential function as in \cref{fig:phaseflip_evolution}. (c) Slope $y$ is obtained from a linear fit $\Gamma_{\mathrm{dep}} = x + y \ev{\hat{n}_{\mathrm{ss}}}$ of the lines in (a) and (b).}
    \label{fig:phaseflip}
\end{figure}

A more complete analysis for more combinations of $n$ and $m$ is done in \cref{fig:phaseflip}, where we also compute the dephasing rate for different values of the mean photon number. We find that cat-states survive longer in a resonator with $n = m$. We also find a consistent linear relationship between the dephasing error rate and the mean photon number similar to the cases $n = m$, indicating that states with more photons are more sensitive to noise. However, the slope of the curves differs drastically depending on the relation between the two degrees $n$ and $m$. As can be seen in \cref{fig:phaseflip}(c), we show the slope $y$ obtained from a linear fit $\Gamma_{\mathrm{dep}} = x + y \ev{\hat{n}_{\mathrm{ss}}}$ of the lines in panels (a,b) (as well as from the data obtained for the lines not shown). Essentially, when $n$ and $m$ are coprime, the dephasing error rate increases very fast with the mean photon number with slopes of up to four orders of magnitude larger than the corresponding coherent-state cases. Instead, if $\gcd(n,m)> 1$, we obtain a much smaller scaling comparable to the situations with $n=m$. For instance, for a driving degree of $2$, we have $\Gamma_{\mathrm{dep}} = 2\gamma_1$ for both $m=2$ and $m=4$. This result is compatible with the theoretical scaling of the dephasing rate \cite{lescanne2020exponential}. Hence, even though we are in the weak-symmetry regime where cat states are not the steady states of the system, during the metastability window, an effective decoherence-free subspace appears in the metastable manifold that freezes the dynamics until the decay to the single steady state \cite{kouzelis2020dissipative}. The fact that only the linear dissipative term mixes the symmetry blocks allows us to maintain coherent superpositions for a longer time. Another example of this phenomenon, which is not shown in \cref{fig:phaseflip}, is the case of $n=3$ and $m= 6$. There, we obtain a slope of the dephasing rate equal to $1.84 \pm 0.03$ which is close to the scaling found for $n=m=3$.

The most notable situation is when $n = 4$ and $m = 6$ which has $\gcd(4,6) = 2$. Thus, the corresponding strong-symmetry case ($\gamma_1 = 0$) has only two steady states consisting of a combination of even and odd four-cat states. Consequently, if we were to directly compute the dephasing rate for the four four-cat states, we would see a very fast decay of these states to one of the two stable steady states. The chosen state depends on the symmetry eigenvalue of the initial one, that is, the four-cat states with eigenvalues $0$ and $2$ ($1$ and $3$) converge to the steady state with even (odd) parity. In the weak-symmetry regime ($\gamma_1 > 0$), the number of metastable states depends on the mean photon number $\ev{\opn}_{\mathrm{ss}}$:  two for small values coinciding with the strong-symmetry steady states and four for large values corresponding to the lobes. Hence, this resonator, for the particular values of $\ev{\opn}_{\mathrm{ss}}$ considered, is much more useful to store two-dimensional qubits. Indeed, in \cref{fig:phaseflip}(c) the slope of the dephasing error rate corresponds to the storage of even- and odd-parity states. For more details on this particular case, see \cref{apx:4n6m}.

\section{Quantum Associative Memory}\label{sec:qam}

In our previous work \cite{labay2022memory} we demonstrated the applicability of these oscillators in pattern classification within the framework of quantum associative memory focusing on the balanced configuration $n=m$. By leveraging the metastable phase, where the lobes function as attractors of the system dynamics, we successfully discriminated initial states into $n$ coherent states, which served as the memories for storing information. Building upon this approach, we now extend it to incorporate squeezed states, enabling the encoding of information within the four degrees of freedom of a squeezed coherent state \cref{eq:squeezed_state}. The encoding leads to a state that most closely resembles one of the $n$ memories in the metastable phase $\bar{k}$. During the metastable phase, the system dynamics will converge with a high probability to the desired state if the initial separation from the other lobes is negligible. Then a phase-shifted measurement for squeezed states allows extraction of the lobe $k$ to which it has converged to assert the probability $P[\bar{k} = k]$ that it went to the correct memory. 

For our numerical simulations, we consider an initial squeezed coherent state $\ket{\beta,\zeta}$, where $\abs{\beta}^2/\ev{\hat{n}}_{\mathrm{ss}} \in [1/2, 2]$, $\abs{\zeta} \in [0, 1]$, and the phases are randomly chosen from the intervals $[0,2\pi]$ and $[0,\pi]$ for each respective complex value. To determine the closest memory lobe, we calculate $\bar{k} = \mathrm{arg\, min}_{k=1,\dots,n} \norm{\mu_k - \op{\alpha,\zeta}}$, where $\mu_k$ represents the $k$th memory state.

Using the squeezed coherent state as the initial state, we perform a Monte Carlo simulation based on the master equation \ref{eq:pp_me_nm}. The simulation is run for a time long enough to ensure the state penetrates into the metastable transient. At this point, we measure the state using the positive-operator-valued measure (POVM) $\inbrc{ \Pi_j = \op{\alpha_j,\xi_j} }_{j=1}^n$ and $\Pi_? = \Id - \sum_j \Pi_j$ that is used for phase-shifted squeezed state discrimination and represents a natural extension to the POVM used in \cite{labay2022memory}. 

The success probability of correctly identifying the lobe for a single trajectory is given by
\begin{equation}
    P[\bar{k} | \tau_n < t < \tau_2] = \frac{1}{\tau_2 - \tau_n} \int_{\tau_n}^{\tau_2}dt \tr[\Pi_{\bar{k}} \rho(t)]\ ,
\end{equation}
where $\rho(t)$ is the density matrix at time $t$. To obtain reliable statistics, we repeat the entire procedure for different initial states, generating 500 realizations, and calculate the average success probability over these realizations.

\begin{figure*}
    \centering
    \includegraphics[width=\linewidth]{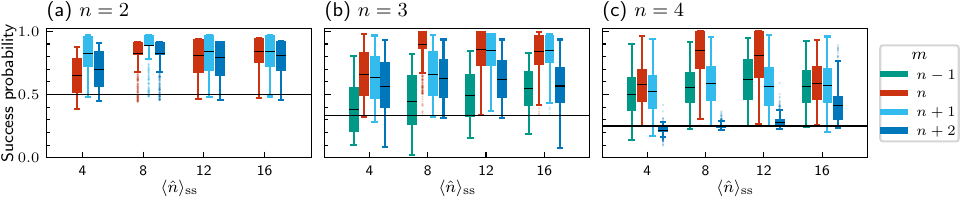}
    \caption{(a)-(c) Success probability for measuring the correct lobe starting from an initial squeezed-coherent state with random amplitude and squeezing parameter for (a) $n=2$, (b) $n=3$ and (c) $n=4$. Each box is obtained from an average of $500$ Monte Carlo trajectories with different initial states. The lower and upper sides of the box correspond to the first and third quartiles, respectively and the inner line denotes the mean probability of success. The position of the whiskers is $1.5$ of the interquartile range, points lying outside this range are shown as dots. The horizontal black line corresponds to the success probability of randomly guessing the lobe, i.e. $1/n$.}
    \label{fig:success_probability}
\end{figure*}

In \cref{fig:success_probability} we show the success probability for an oscillator with mean photon number $\ev{\hat{n}} = 4,8,12$ and $16$. The error bars correspond to the standard deviation of the success probability that is averaged for the 500 trajectories. We include a horizontal black line that represents the success probability one would obtain by randomly guessing the lobe, that is, $1 / n$. In other words, the state is a statistical mixture of all lobes [see \cref{eq:steadystate_weak}]. 

In general, we can see that the success probability tends to increase as the mean number of photons increases. This is expected since the discrimination between the lobes improves the further they are from the center and each other. Moreover, in the large-amplitude regime, the metastable states are well approximated by squeezed states as we saw in \cref{sec:squeezing}. Thus, the POVM used for state discrimination of squeezed states is optimal. 

Another characteristic is that patterns made of squeezed states ($n\neq m$) have a lower success probability than coherent states ($n=m$) except for $n = 2$ where the cases $m > n$ outperform the latter. This is the same behavior we saw previously in \cref{sec:metastability} and \cref{sec:memory}. These two cases with $m=3$ and $m = 4$ had a longer metastable phase which led to longer bit-flip times. In this case, the squeezed lobes achieve a success probability near one for $\ev{\hat{n}} \ge 8$ with the coherent state results slightly below them. However, this trend changes when considering different values of $n$. \Cref{fig:success_probability}(b) reveals that when $m = n + 1$ (amplitude-squeezed states), a higher success probability is obtained compared to other squeezing cases. On the other hand, \cref{fig:success_probability}(c) shows that the highest success probability (excluding the $n = m$ scenario) is observed when $m = n - 1$ (phase-squeezed states). This discrepancy arises because amplitude-squeezed lobes exhibit a smaller distance between them when $n = 4$ compared to the case of $n = 3$, as more states must fit within a fixed photon number. The least favorable scenario occurs in $(4, 6)$, where the success probability is comparable to random guessing because the lobes are not the metastable states\footnote{As explained in \cref{apx:4n6m}, the four lobes are not the metastable phases so a different encoding would be needed to encode only two patterns in the even and odd parity states spanning the metastable manifold.}. Also, when the driving and dissipation degrees are $3$ and $2$ respectively, the lobes are not well characterized by squeezed states unless $\ev{\hat{n}} \ge 16$ (see \cref{sec:squeezing}) which reduces the success probability for smaller mean photon number. Nevertheless, the highest success probability is reached for coherent states with three and four patterns. 

It should be noted that the cut-off set for the amplitude and squeezing of the initial states also affects the performance of the associative memory. In this study, we have limited the minimum amplitude of the lobes setting it to $\sqrt{\ev{\hat{n}} / 2}$. This criterion serves to exclude from the analysis states which are equally spaced from all the lobes and lead to a success probability of $1/n$. Nevertheless, one can note in \cref{fig:success_probability} that whiskers and outliers points arrive on several occasions at the random guessing success probability. Similarly, the maximum squeezing has been limited to $1$ ($\SI{-8.68}{dB}$) which already allows for the application of quantum key distribution protocols \cite{preskill2001qkd}  but may negatively affect the performance depending on its direction. For instance, an amplitude-squeezed state with a large squeezing parameter might overlap two lobes.

We recall that in associative memories, the storage capacity is defined as the number of memories stored in the system over its dimension. This quantity has a classical bound of $\alpha_c = 0.138$ for an all-to-all network of binary neurons (commonly known as a Hopfield neural network) \cite{hopfield1982neural,amit_1989}. This limit has not yet been surpassed by the quantum version of the Hopfield network made of spin-$1/2$ units \cite{enhancing2021marsh,fiorelli2023storagecapacity} although quantum systems promise to store an exponentially large number of patterns \cite{lewenstein2021storagecapacity}. In \ccite{labay2022memory}, we showed that by optimizing the amplitude of the lobes one can overcome the classical bound and reduce the dimension of the Hilbert space needed to store a given number of patterns. In the presence of squeezed states, the storage capacity can be enhanced for amplitude-squeezed states as they can be described using a smaller Hilbert space. However, the capacity to distinguish the patterns is highly affected by small mean photon numbers. Hence, the combination of the two factors makes coherent-state storage more optimal in terms of storage capacity for $n > 2$.

\section{Conclusions}\label{sec:discussion}
In this work, we studied several dynamical properties of a quantum oscillator with driving and dissipative terms that exchange photons with the environment in packets of $n$ and $m$ particles. A rich scenario of dynamical behaviors was reported in connection with different symmetries and spectral features of the Liouvillian. This led to the possibility of obtaining steady states with symmetrically phase-distributed lobes that can be characterized as squeezed-coherent states, especially for high driving strength and small nonlinear dissipation rate. We have seen that a higher driving degree leads to phase-squeezed states while a higher dissipation degree leads to amplitude-squeezed lobes.

In terms of the metastable phase, when linear damping is counted, we have seen that the lobes are well approximated by squeezed states when the driving degree is equal to the dissipation degree. This is the case for coherent states ($n=m$) and phase-squeezed states ($n=m+1$). In the case of amplitude-squeezed states ($n=m-1$), the lobes are not well approximated by squeezed states unless the mean photon number is large enough. This is because the lobes are closer to each other and the squeezing parameter must be large enough to distinguish them.

We have analyzed two applications of the oscillator well suited to the metastable regime. The first one is the storage of quantum states where we have characterized the relaxation and dephasing times of qudits stored in the metastable phases. We have seen that the relaxation time for squeezed states is longer than for coherent states for $n = 2$. This is because the lobes are further apart from each other and the spectral gap is larger. Moreover, while the phase flip error rate is expected to grow linearly with the mean photon number, we have obtained an exponential scaling when the two non-linear degrees are coprime. Conversely, the dephasing error rate scaling when $\gcd(n,m) > 1$ remains linear even in unbalanced situations where the scale factor with the mean photon number can be improved in some cases as compared to balanced models ($n=m$). Hence, coherences between states disappear at the same or a smaller rate than for coherent states.

The second application is quantum associative memory where the same system in the metastable regime is used to recognize (squeezed-states) memories. Following from the results of \ccite{labay2022memory}, we analyzed the possibility of storing and retrieving genuine quantum states and computed the success probability of pattern discrimination for the different driving and dissipative degrees. We have seen that in general amplitude-squeezed states are a better option than phase-squeezed states to store the patterns. Nevertheless, patterns encoded in a coherent state allow attaining a high success probability for a smaller mean photon number (system size). This shows the possibility of using such memories for practical purposes. For instance, in discrete modulated continuous-variable quantum key distribution \cite{lin2019qkd,huang2014qkd}, the information encoded in squeezed coherent states can be distorted due to the transmission channel \cite{xuan2009experiment}. Quantum associative memory can be used to retrieve the original information.

Our primary emphasis centered on exploring how squeezing could potentially be employed to improve the capabilities of our driven-dissipative oscillator in storing and retrieving information problems.  The resonator, however, has the potential to be used for more tasks including quantum error correction \cite{Terhal_2020,minganti2022squeezedcat,Gertler2021} or holonomic quantum control \cite{albert2016holonomic}.

Finally, we exhaustively analyzed some properties of the oscillator necessary to realize the aforementioned applications. Despite that, the rich dynamical features present go far beyond the ones presented. The presence of dissipative phase transitions found in the balanced model \cite{minganti2023dissipative}, exceptional points in the Liouvillian spectrum \cite{minganti2019ep,cabot2021metastable} or symmetry breaking \cite{cabot2023nonequilibrium,lieu2020symmetryQEC} might be of further interest in the generalized model.

\begin{acknowledgments}
We acknowledge the Spanish State Research Agency, through the Mar\'ia de Maeztu project CEX2021-001164-M funded by the MCIN/AEI/10.13039/501100011033, through the QUARESC project (PID2019-109094GB-C21/AEI/10.13039/501100011033) and through the COQUSY Projects No. PID2022-140506NB-C21 and No. PID2022-140506NB-C22 funded by MCIN/AEI/10.13039/501100011033, Ministry for Digital Transformation and of Civil Service of the Spanish Government through the QUANTUM ENIA project call - Quantum Spain project, and by the European Union through the Recovery, Transformation and Resilience Plan - NextGenerationEU within the framework of the Digital Spain 2026 Agenda. The CSIC Interdisciplinary Thematic Platform (PTI+) on Quantum Technologies in Spain (QTEP+) is also acknowledged. GLG is funded by the Spanish  Ministerio de Educaci\'on y Formaci\'on Profesional/Ministerio de Universidades and co-funded by the University of the Balearic Islands through the Beatriz Galindo program (BG20/00085). ALM is funded by the University of the Balearic Islands through the project BGRH-UIB-2021.
 \end{acknowledgments}

\appendix
\renewcommand{\thefigure}{A\arabic{figure}}
\setcounter{figure}{0}

\section{Mean field}\label{apx:mean_field}

The evolution of the expectation value of the operator $\opa$ can be used to characterize the mean-field dynamics. For that, we compute from the master equation \cref{eq:pp_me_nm}  $\dot{\alpha} = \tr(\opa \dot{\rho})$ and approximate $\ev{\opa^x (\opad)^y} \sim \ev{\opa}^x\ev{\opad}^y$. This leads to the equation \cite{labay2022memory}
\begin{equation} \label{eq:pp_mf_a}
    \dot{\alpha} = - \frac{\gamma_1}{2}\alpha -i \Delta \alpha - n \eta (\alpha^*)^{n - 1}e^{-i n\theta} - \frac{m}{2}\gamma_m \abs{\alpha}^{2(m - 1)} \alpha\ ,
\end{equation}
which becomes exact in the thermodynamic limit of a large number of excitations ($\abs{\alpha}^2 \to \infty$) \cite{minganti2023dissipative}.

The roots of \cref{eq:pp_mf_a} determine the fixed points of the system. Defining $\alpha = R e^{i\phi}$, we obtain the mean field amplitude 
\begin{equation} \label{eq:mf_amplitude}
    R^{2m - n} = \frac{2 n \eta_n}{m \gamma_m} \ .
\end{equation}
of the $n$ symmetrically distributed lobes forming the steady state. This equation is only valid for $2m > n$ which will be the cases studied in this paper, and is only valid for large amplitudes $R$. In these conditions, the $n$ fixed points are symmetrically distributed with angles $\theta_j = (2j + 1)\pi / n$ where $j = 1,\dots,n$.

A general solution for the amplitude $R$ does not exist for all powers $n$ and $m$ but, in some cases, we can get a more accurate description by fixing one of the two exponents. Concretely, for $n = 2$ and $m > 1$ we have\begin{equation} \label{eq:mf_amplitude2}
    R^{2 m - 2} = \frac{2}{m\gamma_m} \insqr{\sqrt{(2\eta)^2 + \Delta^2} - \frac{\gamma_1}{2}}
\end{equation}
and for $m = n - 1$ we get
\begin{widetext}
\begin{equation} \label{eq:mf_amplituden1}
    R^{2n - 4} =\frac{1}{[(n-1)\gamma_m]^2} \inbrc{ \insqr{2(n\eta)^2 - (n-1)\gamma_m\gamma_1} + \sqrt{\insqr{2(n\eta)^2-(n-1)\gamma_m\gamma_1}^2 - [(n-1)\gamma_m]^2 \insqr{\gamma_1^2 + (2\Delta)^2}}}\quad .
\end{equation}
This expression is especially useful in the case $(3,2)$ where the approximation in \cref{eq:mf_amplitude} fails in the regimes considered in this work.

The stability of the fixed points can be analyzed easily for $\gamma_1 = \Delta = 0$. In this case, the Jacobian matrix of the system is
\begin{equation}
    J[R,\phi] = \mqty(-\half \gamma_1 -\frac{m(2m-1)}{2}\gamma_m R^{2m - 2} -n (n -1)\eta R^{n-2}\cos n\phi & n^2\eta R^{n-1}\sin n\phi \\ n(n-2) \eta R^{n-3}\sin n\phi & n \eta R^{n-2}\cos n\phi )\ .
\end{equation}
\end{widetext}
Substituting \cref{eq:mf_amplitude} and $n\phi = (2j+1)\pi$ gives 
\begin{equation}
    J[R_{fp},\phi_{fp}] = \mqty(-\half \gamma_1 + n\eta R^{n - 2}(n - 2m) & 0 \\ 0 & -n\eta R^{n-2} )\ ,
\end{equation}
so if $n - 2m < 0$ the eigenvalues are negative and lead to $n$ stable fix points. This is the case for all pairs of $(n,m)$ presented in this paper.

\begin{figure}
    \centering
    \includegraphics[width=\linewidth]{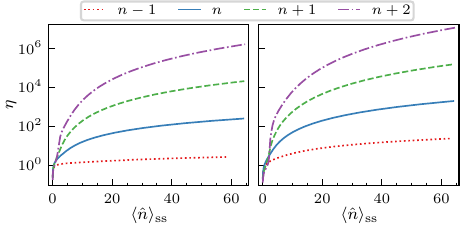}
    \caption{Strength of the driving needed to generate the lobes over their mean photon number for (a) $n=3$ and (b) $n=4$. }
    \label{fig:driving_photon_number}
\end{figure}

It is important to note that the driving strength $\eta$ necessary to achieve a steady state with mean photon number $\ev{\opn}_{\mathrm{ss}}$ differs greatly for different values of the nonlinear degrees $(n, m)$. In general, as can be seen in \cref{fig:driving_photon_number}, a two-order of magnitude increase in $\eta$ is needed for each additional photon lost in the nonlinear dissipation degree. The effect is similar to increasing $\gamma_m$ [see \cref{eq:mf_amplitude}], photons are lost more rapidly and so a higher $\eta$ is necessary to stabilize the system \cite{sonar2018squeezing}.

Given that the value of $\eta$ changes abruptly for the different combinations of $(n,m)$, we use $\ev{\opn}_{\mathrm{ss}}$ as a suitable parametrization to show the results. This allows us to use the same scale for all nonlinear degrees, but we must note that the actual parameter being changed is $\eta$.

\section{Four-photon driving and six-photon dissipation}\label{apx:4n6m}

The oscillator with $n = 4$ and $m = 6$ behaves differently from the other scenarios considered with four lobes. The difference is explained by the relation between the two nonlinear degrees which have $p = \gcd(4,6) = 2$ and so $p \neq n$. Hence, in the absence of linear dissipation, a strong symmetry arises with only two steady states of even and odd parity, in contrast, for instance, to the $n=m=4$ case where there are four steady states corresponding to four-cat states each having a different symmetry eigenvalue. 

\begin{figure}
    \centering
    \includegraphics[width=\linewidth]{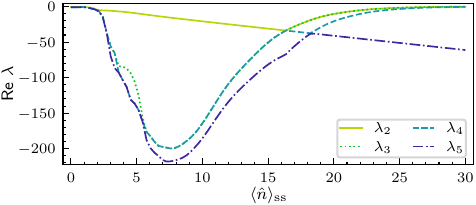}
    \caption{Evolution of the first Liouvillian eigenvalues for increasing mean photon number of the steady state. The first eigenvalue is omitted as it vanishes exactly. The parameters: $\gamma_1 = 1.0$, $\gamma_4 = 0.2$ and $\Delta = 0.4$.}
    \label{fig:gap_ratio_46}
\end{figure}

Notably, this also has consequences in the weak symmetry case. Two distinct regimes can be appreciated in \cref{fig:gap_ratio_46} from the distribution of the Liouvillian spectrum. Initially, for a small mean photon number, the fourth and fifth eigenvalues are close to each other and the largest eigenvalue separation occurs between $\lambda_2$ and $\lambda_3$. Then, at around $\ev{\opn}\approx 20$, a high order exceptional point \cite{heiss2012ep,minganti2019ep} occurs between the eigenvalues $\lambda_2$, $\lambda_3$ and $\lambda_4$ that makes the first two complex conjugate ($\lambda_2 = \lambda_3^*$) and the last one real. Beyond the exceptional point, the separation between the fourth and fifth eigenvalues becomes larger indicating that a four-dimensional metastable phase may arise for a very large photon number. This makes sense since for small $\ev{\opn}$ the driving strength $\eta$ and the nonlinear dissipation rate $\gamma_m$ are comparable so the symmetry that dominates is $\mathds{Z}_2$. For larger $\ev{\opn}$, the driving strength is much larger than the nonlinear dissipation rate so the symmetry that dominates is $\mathds{Z}_4$. In all cases, however, the spectral separation is small compared to the other cases with $n = m$.

As we work in the regime of small mean photon number $\ev{\opn} < 20$, only two metastable states are present. To characterize these states we start by defining $\{ \ket{\psi_j} \}_{j=0}^3$ as the squeezed-coherent states describing the four symmetrically distributed lobes. Then, the four four-cat states can be written as
\begin{eqs}
    \ket{\pi_0} &= \half \inpar{ \ket{\psi_0} + \ket{\psi_1} + \ket{\psi_2} + \ket{\psi_3} } \\
    \ket{\pi_1} &= \half \inpar{ \ket{\psi_0} - i\ket{\psi_1} - \ket{\psi_2} + i\ket{\psi_3} } \\
    \ket{\pi_2} &= \half \inpar{ \ket{\psi_0} - \ket{\psi_1} + \ket{\psi_2} - \ket{\psi_3} } \\
    \ket{\pi_3} &= \half \inpar{ \ket{\psi_0} + i\ket{\psi_1} - \ket{\psi_2} - i\ket{\psi_3} } 
\end{eqs}
Each of these states $\{ \ket{\pi_j} \}$ has only Fock levels $\ket{a}$ with $a \mod 4 = j$. Combining the two four-cat states with even and odd parity we obtain 
\begin{eqs}
    \ket{C_\pm^{\text{even}}} &= \frac{1}{\sqrt{2}}\inpar{ \ket{\psi_0} \pm \ket{\psi_2} } \\
    \ket{C_\pm^{\text{odd}}} &= \frac{1}{\sqrt{2}}\inpar{ \ket{\psi_1} \pm \ket{\psi_3} }
\end{eqs}
from which we can construct the two metastable states
\begin{eqs}[eqs:46_ems]
    \mu_0 &= \half \inpar{\op{C_+^{\text{even}}} + \op{C_+^{\text{odd}}}} \\
    \mu_1 &= \half \inpar{\op{C_-^{\text{even}}} + \op{C_-^{\text{odd}}}}
\end{eqs}
where $\mu_0$ ($\mu_1$) has even (odd) parity and corresponds to the steady states of the system in the limit $\gamma_1 \to 0$.

\begin{figure}
    \centering
    \includegraphics[width=\linewidth]{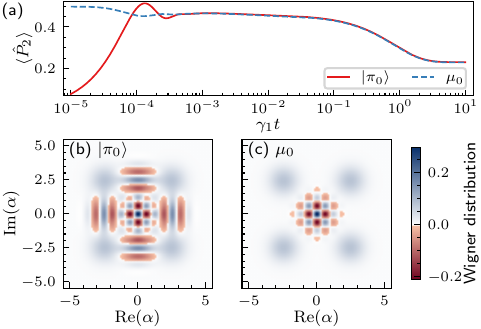}
    \caption{Time evolution of the expectation value of the parity operator $\hat{P}_2$ containing only Fock states $\ket{a}$ where $a\mod 4 = 2$. The Wigner distribution of the initial states can be seen for (b) the parity 0 cat-state $\ket{\pi_0}$ and (c) the even parity metastable state $\mu_0$.}
    \label{fig:46_traj}
\end{figure}

\Cref{fig:46_traj} shows the time evolution of two initial states: $\ket{\pi_0}$ and $\mu_0$. The former is an even parity four-cat state containing only Fock states with a photon number multiple of $4$. The latter, instead, is the metastable state with even parity (it is spanned by all even Fock states). In \cref{fig:46_traj}(a) we plot the expectation value of the parity operator $\hat{P}_2 = \sum_a \op*{2 a + 2}$. This operator allows us to see that the cat states (which are not metastable states) lead to one of the two metastable states specified in \cref{eqs:46_ems}. More specifically, initially the state $\ket{\pi_0}$ has no component on the subspace spanned by the Fock states with modulus 2 but rapidly, the state converges to $\langle \hat{P}_2 \rangle = 1/2$, which we would expect for a state that has all the even modes populated. On the other hand, the metastable state $\mu_0$ remains in the same state for a longer time. Hence, the parity of the initial cat state determines to which metastable state it converges.

This has consequences in both applications we studied: storage and quantum associative memory. First, in terms of storage of quantum states, the memory should be regarded as a two-dimensional system where the two metastable states are the computational basis states. Thus, the relaxation and dephasing times should be calculated accordingly. Second, in terms of quantum associative memory, the number of patterns that can be efficiently stored is only 2. Trying to store four patterns in the squeezed lobes leads to a success probability close to $1/4$ because the states rapidly decay to the manifold $\{ \mu_0, \mu_1 \}$ and the posterior state discrimination is not capable of distinguishing between the two states. Hence, a different encoding would be needed, for instance, knowing that Fock states with even (odd) parity converge to metastable states with even (odd) parity. We can use this property to encode the initial states which would allow us to use the system for pattern discrimination of two memories.

\bibliography{references.bib}

\end{document}